\documentclass[12pt]{article}
\usepackage{amsmath}
\usepackage{graphicx}
\usepackage[numbers]{natbib}
\usepackage{algorithm2e}
\usepackage{latexsym}
\usepackage{multirow}
\usepackage{amsfonts}
\usepackage{amssymb}
\usepackage{color}
\usepackage{enumerate}
\usepackage{geometry}
\usepackage{lscape}
\usepackage{url}
\usepackage{float}

\restylefloat{table}
\restylefloat{figure}

\usepackage{helvet}

\def\bX{\mathbf{X}}

\def\bt{\mathbf{t}}

\def\bx{\mathbf{x}}

\makeatletter
\def\url@leostyle{%
  \@ifundefined{selectfont}{\def\UrlFont{\sf}}{\def\UrlFont{\small\ttfamily}}}
\makeatother
\newcommand{\Keywords}[1]{\par\noindent
{\small{\em Keywords\/}: #1}}
\begin{document}
\title{Parametrization of white matter manifold-like structures using principal surfaces}
\author{Chen Yue$^{a}$, Vadim Zipunnikov$^{a}$, Pierre-Louis Bazin$^{c}$, \\ Dzung Pham$^{b}$, Daniel Reich$^b$, Ciprian Crainiceanu$^{a}$, Brian Caffo$^a$\\
\footnotesize $^{a}$  Department of Biostatistics, Johns Hopkins University, Baltimore, MD, 21205\\
\footnotesize $^{b}$ Department of Radiology and Imaging Sciences, National Institute of Health, Bethesda, MD 20892\\
\footnotesize $^{c}$ Department of Neurophysics, Max Planck Institute, Leipzig, Germany, 04103
}
\date{}
\maketitle

\begin{abstract}
In this manuscript, we are concerned with data generated from a diffusion tensor imaging (DTI) experiment. The goal is to parameterize manifold-like white matter tracts, such as the corpus callosum, using principal surfaces. We approach the problem by finding a geometrically motivated surface-based representation of the corpus callosum and visualize the fractional anisotropy (FA) values projected onto the surface; the method applies to any other diffusion summary as well as to other white matter tracts. We provide an algorithm that 1) constructs the principal surface of a corpus callosum; 2) flattens the surface into a parametric 2D map; 3) projects associated FA values on the map. The algorithm was applied to a longitudinal study containing 466 diffusion tensor images of 176 multiple sclerosis (MS) patients observed at multiple visits. For each subject and visit the study contains a registered DTI scan of the corpus callosum at roughly 20,000 voxels. Extensive simulation studies demonstrate fast convergence and robust performance of the algorithm under a variety of challenging scenarios.

\end{abstract}
\newpage
\section{Introduction}
This research is motivated by the need to establish a parametric description of the corpus callosum structure, a neuronal structure with a curved shape. The corpus callosum is a wide, flat and curved bundle of neural fibers beneath the cortex in the eutherian brain at the longitudinal fissure. It connects the left and right cerebral hemispheres and facilitates interhemispheric communication. As the largest white matter structure in the human brain, it consists of 200 to 250 million contralateral axonal projections.\

\begin{figure}[H]
\caption[Corpus callosum data illustration]{\footnotesize Left panel: 3D-rendering of corpus callosum. Right panel: horizontal, sagittal, and coronal slices. Views: R=Right, L=Left, S=Superior, I=Interior, A=Anterior, P=Posterior. Produced in 3DSlicer (4.1.1)}
\label{fig.corpus}
\begin{minipage}[b]{0.45\linewidth}
\centering
\scalebox{0.4}{\includegraphics{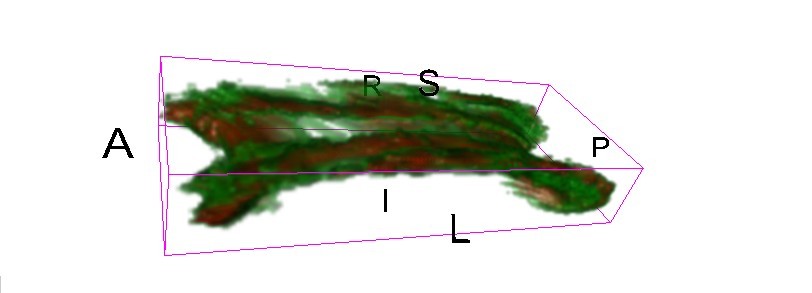}}
\end{minipage}
\begin{minipage}[b]{0.45\linewidth}
\scalebox{0.4}{\includegraphics{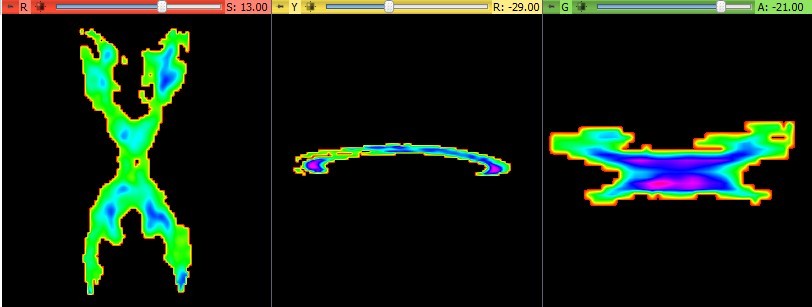}}
\end{minipage}
\end{figure}

The left panel in Figure \ref{fig.corpus} displays that the corpus callosum appears as a two dimensional manifold in its principal structure. It is curved towards the inferior part of the brain on both the anterior and posterior sides. A different perspective is shown in the middle panel on the right. Though the corpus callosum lies in a three dimensional space, its key structure is intuitively that of a ``carpet" that lies in a two dimensional manifold. Therefore, dimension reduction techniques may provide strong data compression along with novel visualization and parametrization approaches that could be easy to use in practice.

To achieve this, the first step is to estimate the center surface of the corpus callosum. The second step is to obtain the projection of each data point; these projections can then be used to map the diffusion properties of the corpus callosum onto the 2D manifold.\

The focus of our work is on principal surfaces for dimension reduction. Before we describe them, we outline other potential methods. Principal component analysis (PCA) is one of the most useful tools for dimension reduction. PCA finds directions (vectors) in the space that explain the largest variability of the data, while constraining directions of variation to be orthogonal.\

\begin{figure}[H]
\caption[Principal curves and surfaces]{\footnotesize Left panel is a illustration of different dimension reduction methods. The blue points are the original data points, the dot-dashed green line is the first principal component, the dashed black curve is the spline fitting and the solid red curve is one of the principal curves. Middle panel shows the difference between the principal curve and the regression method: top panel shows that the principal curve minimizes the orthogonal distance, bottom panel shows that the spline regression minimize the distance in y axis. Both panels in the middle use the same dataset. Right panel illustrates several different principal curves for one dataset.}
\label{fig.def}
\begin{minipage}[b]{0.45\linewidth}
\centering
\scalebox{0.37}{\includegraphics{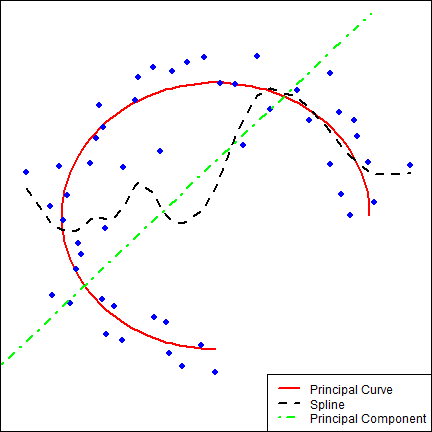}}
\end{minipage}
\hspace{-1.2cm}
\begin{minipage}[b]{0.45\linewidth}
\centering
\scalebox{0.4}{\includegraphics{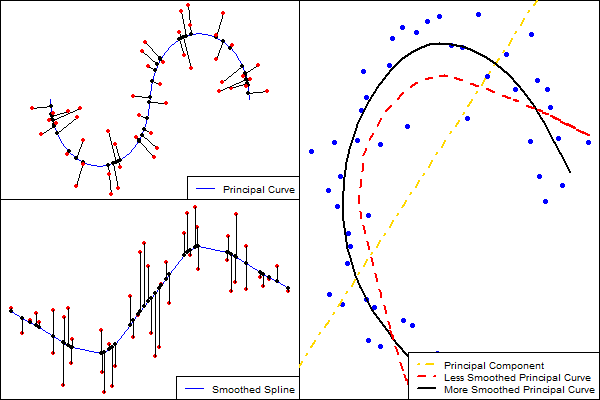}}
\end{minipage}
\end{figure}

Though PCA is a widely used strategy for dimension reduction, it can lead to over simplification. As shown in the left panel in Figure \ref{fig.def}, the underlying true curve that generates the data is part of a circle while the first principal component is a line. Direct applications of splines, wavelets and related regression methods cannot be done as they require the mean function of the data to be in a functional form; that is, there is only one $y$ for every $x$. This is clearly violated in the corpus callosum example and many other white matter manifolds of interest. \citeauthor{palus1992singular} \cite{palus1992singular} illustrated the pitfalls and precautions when applying linear PCA in non-linear settings. In our case, since the corpus callosum clearly is a nonlinear structure, PCA is not a viable candidate for dimension reduction.\

Many non-linear methods have been proposed for fitting non-linear data structures. As an example, \cite{gnanadesikan1977methods} proposed a non-linear extension of PCA. The core idea is to include product combinations of the variables in the data matrix. Another useful tool -- the self-organizing map (SOM) -- was proposed in \cite{kohunen1990theself}, \cite{kohonen1982self-organize}. SOMs are unsupervised learning procedures which are used to discover structure in the data. Other nonlinear method methods such as non-linear principal component analysis (NLPCA) \cite{kramer1992autoassociative}, \cite{kramer1991nonlinear} and principal geodesic analysis (PGA) \cite{fletcher2004medical}, \cite{fletcher2004statistical} can also fit non-linear data structures.

An important concept in doing non-linear data compression is principal surfaces \cite{hastie1984principal}, \cite{hastie1989principal}. Principal surfaces are manifolds that pass through the middle of the data. The surfaces are fit via nonparametric low-dimensional manifolds that minimize the orthogonal distance from the data to themselves. Principal surfaces satisfy a {\em self consistency} condition, in that they are the conditional expectation (local average) of the data. The middle panel of Figure \ref{fig.def} shows the difference between the principal curves (surfaces) and regression. It highlights that the principal curve minimizes the sum of orthogonal distances while a spline model fit tries to minimize the sum of distances parallel to the y axis.\

The principal surface is not unique. For example, it has been proven \cite{hastie1984principal}, \cite{hastie1989principal} that all the principal components of a dataset are self-consistent. Other factors may also impact the results of fitting principal curves and surfaces. Using another example, different smoothness and lengths also lead to different principal curves. The original algorithm proposed in \cite{hastie1989principal} used an iterated algorithm with local planar smoother to achieve the principal surfaces. Motivated by multivariate adaptive regression splines (MARS), adaptive principal surface is proposed later in \cite{leblanc1994adaptive}. \citeauthor{dong1996nonlinear} \cite{dong1996nonlinear} combines principal curves and neural networks. The authors construct a principal curve via a two-step neural network where the nonlinear ``layer" is the construction of principal curve. \citeauthor{einbech2010data} \cite{einbech2010data} proposed an algorithm for fitting local principal surfaces. The surface is connected via triangles and the main work is then to find adjacent vertices. \citeauthor{goldsmith2011nonlinear}\cite{goldsmith2011nonlinear} proposed a tube fitting algorithm based on principal curve method. Other algorithms have been proposed recently, such as fitting principal surfaces via kernel map manifold (KMM methods) \cite{gerber2009dimensionality} and constructing local principal curves and surfaces using subspace constraint mean shift (SCMS) \cite{ozertem2011locally}.\

Some local search methods though could fit the principal surface, may not have equally spaced parametrization on the surface. This will definitely lead to poor visualization of the flattened surface. In addition, some previous work can only achieve locally planar surfaces while for corpus callosum and other white matter tracts, a surface with certain degree of smoothness would be more reasonable. Therefore, we develop a method of obtaining a smoothed principal surface of the data with better parametrization on the surface. The rest of the manuscript is laid out as follows: In section 2, the principal surface concept will be introduced, and its corresponding algorithm will be shown. We will show how the principal surface algorithm works on simulated data in section 3. The algorithm then is applied to corpus callosum data and the FA maps are obtained in section 4. We will conclude the whole paper in section 5.\

\section{Methods}

\subsection{Principal Surfaces}
Let $\mathbf{x}_i=(x_{i1},x_{i2},x_{i3})^T,\ i=1,\dots, I$ be the data points in three dimensional space, $\mathcal{R}^3$ following an underlying distribution. $\mathbf{t}_i=(t_{i1},t_{i2})^T$ be corresponding parametrization points in two dimensional space, $\mathcal{R}^2$. In addition, we will require (without loss of generality) that the 2D coordinate space be the unit square $[0,1]\times[0,1]\subset \mathcal{R}^2$. We define $f$ as the smooth principal surface function $f: \mathbf{t}_i\mapsto f(\mathbf{t}_i)$ that maps from $\mathcal{R}^2$ to $\mathcal{R}^3$. The principal surface function satisfies the {\em self-consistency} condition:
\begin{equation}
E(\bX|\lambda_f(\bX)=\bt)=f(\bt)\ \ \text{for all }\bt,
\end{equation}
where $\lambda_f(\bx)=\sup_{\mathbf{t}}\big\{\mathbf{t}: \|\mathbf{x}-f(\mathbf{t})\|=\inf_{\mathbf{\mu}}\|\mathbf{x}-f(\mathbf{\mu})\| \big\}$ is the projection function with respect to $f$. The projection function maps a data point on to the closest principal surface point having the largest parametrization. \citeauthor{hastie1989principal} \cite{hastie1989principal} showed that principal curves and surfaces generally exist, though are not unique. We have found that there are two main distinctions between different fitted principal surfaces: the degree of smoothness and the method of parametrization. In most algorithms, these properties will be controlled by the specific smoother being used in the algorithm and its tuning parameters. The details of our specific algorithm will be demonstrated in the next section.\

\subsection{Algorithm}
In this paper, the algorithm used in \cite{hastie1989principal} will be modified. This algorithm allows us to find the surface coordinate for each data point ($\mathbf{t}_i, i=1,\dots, I$), which we will use later to create parametric summaries. However, the original principal surface algorithm can only yield surfaces which are locally flattened. Therefore, instead of local planar smoothers, we employ thin-plate splines (TPS) for fitting the surface. Thin-plate splines were proposed by \cite{duchon1977splines} and are now widely used for bivariate smoothing. The TPS penalize the least squares error by a high-order derivative term in order to achieve a desired degree of smoothness. \citeauthor{wood2003thin} \cite{wood2003thin} improved the computational efficiency when fitting TPS by using an optimal approximating basis that we employ.\

{\bf Preprocessing.} Recall the notation we defined in section 2.2, let $\mathbf{X}=[\mathbf{x}_1, \dots, \mathbf{x}_I]^T$ be the $I\times 3$ matrix that contains the 3D coordinates of the dataset. We assumed that the data are centered around the origin. Principal component decomposition is then applied. Let $\mathbf{X}=\mathbf{U}\mathbf{\Sigma}\mathbf{V}^T$ be the singular value decomposition of $\mathbf{X}$. Then $\mathbf{U}\tilde{\mathbf{\Sigma}}$ are the first two principal ``scores" of the data matrix, where $\tilde{\mathbf{\Sigma}}$ is a submatrix of $\mathbf{\Sigma}$ containing the first two columns. We then standardized the scores so that both scores are in $[0,\ 1].$ and used them as the initial 2D parametrization.\

\begin{figure}[H]
\caption[Principal surfaces algorithm]{\footnotesize Left panel illustrates the projection step and the conditional expectation step, right panel illustrates the smoothing step}
\label{fig.alg}
\begin{minipage}[b]{0.45\linewidth}
\centering
\scalebox{0.7}{\includegraphics{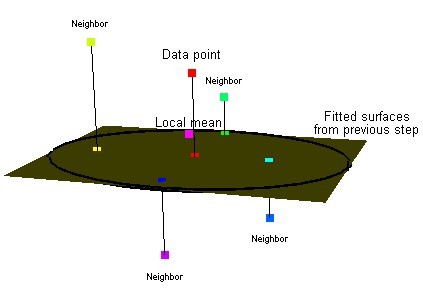}}
\end{minipage}
\hspace{1cm}
\begin{minipage}[b]{0.45\linewidth}
\centering
\scalebox{0.6}{\includegraphics{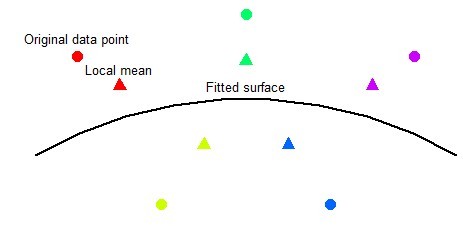}}
\end{minipage}
\end{figure}

{\bf Conditional Expectation.} Suppose $\mathbf{t}_i=(t_{i1},t_{i2})^T$ be the current parametrization of $\mathbf{x}_i$ in two dimensional space, $[0,1]\times[0,1]$. For a specific data point $\bx_0$, with its coordinate $\bt_0$, we choose $r$ as a radius, such that all the other data points with their projection coordinate having less distance from $\bt_0$ than $r$ are considered as the neighbor of $\bx$, see the left panel in Figure \ref{fig.alg}. Let $\mathcal{N}_{\bx_0}$ be the set of all the neighbors of the point $\bx_0$. Then we have $\mathcal{N}_{\bx_0}=\big\{\bx_j :\ \|\bt_{j}-\bt_0\|\le r \big\}$ and we calculate the local weighted average as follows:
\begin{equation*}
\bx^{lm}=\sum_{i}w^{\bx}_i \bx_i\ \ \text{where}\ \ w_i^{\bx}:=\mathbf{1}_{[\mathbf{x}_i\in \mathcal{N}_{\mathbf{x}_0}]}\times\frac{\exp\big\{-\frac{\|\bt_i-\bt\|_2^2}{h}\big\}} {\sum_{j:\ \bx_j\in \mathcal{N}_{\bx_0}}\exp\big\{-\frac{\|\bt_j-\bt\|_2^2}{h}\big\}}.
\end{equation*}

{\bf Smoothing.} Bivariate thin plate splines \cite{wood2003thin} are applied after we obtained these local averages: $\bx_i^{lm},\ i=1,\dots,I$. In this step, we fitted
\begin{equation}
x^{lm}_{id}=f_{d1}(t_{i1})+f_{d2}(t_{i2})+f_{d3}(t_{i1},t_{i2})+\epsilon_{id},\ \ d=1,2,3,
\end{equation}
and obtained a bivariate thin plate spline smoothing of the local average points, which is demonstrated in the right panel of Figure \ref{fig.alg}. Here
\begin{equation}
f(\bt_i)=\left[\begin{array}{l}\hat{f}_{11}(t_{i1})+\hat{f}_{12}(t_{i2})+\hat{f}_{13}(t_{i1},t_{i2})\\
\hat{f}_{21}(t_{i1})+\hat{f}_{22}(t_{i2})+\hat{f}_{23}(t_{i1},t_{i2})\\ \hat{f}_{31}(t_{i1})+\hat{f}_{32}(t_{i2})+\hat{f}_{33}(t_{i1},t_{i2})\end{array}\right]
\end{equation}
is the current principal surface function mapping from the 2D parametrization space to 3D coordinate space. The thin plate splines could be fitted using ``mgcv" package in software {\it R}.

{\bf Projection.} We then projected each data point onto the current principal surface and obtained new 2D parametrization. we used a grid search method to find the projection and only search within the range $[0,1]\times[0,1]$. Therefore there will be some data points being projected onto the boundary, which brings some issues when we further analyzed the 2D parametrization. After this step, we iterated the whole procedure with the new 2D parametrization.\

The whole algorithm is illustrated as follows:\

\begin{algorithm}[H]
\caption{Principal surface algorithm}
\KwIn{Data in 3D coordinate, $\mathbf{X}_{I\times 3}$}
\KwOut{Principal surface function, $f_{\mathcal{R}^2\to \mathcal{R}^3}$, and the 2D parametrization of all data points, $\mathbf{T}_{I\times 2}$}
{\bf Initialization}: De-mean $\mathbf{X}$ for each column\; set the initial 2D parametrization as $\mathbf{T}^{(0)}=\mathbf{U}\tilde{\mathbf{\Sigma}}$\;
Set {\it err}$=1$, i$=1$\;
 \While{($i$ $<$max.iter {\bf and} err$>$thres)}{
  (1). $\mathbf{X}^{lm}\longleftarrow \mathbf{W}\mathbf{X}$, where $w_{ij}=\mathbf{1}_{[\mathbf{x}^{(i)}\in \mathcal{N}_{\mathbf{x}^{(j)}}]}w_{j}^{(\mathbf{x}_i)}$\;
  (2). Fit $\mathbf{X}^{lm}=f(\mathbf{T})+\mathbf{\epsilon}$\;
  (3). $\mathbf{t}^{(new)}_i\longleftarrow arg\min_{\mathbf{t}}\|\mathbf{x}_i-f(\mathbf{t})\|^2$\;
  (4). $err\longleftarrow\|\mathbf{T^{old}}-\mathbf{T^{(new)}}\|_2^2,\ i\longleftarrow i+1$\;
 }
\end{algorithm}

\section{Simulation Results}
\subsection{Simulation Setup}
We illustrate four examples to investigate the algorithm in highly idealized test settings. The number of data points was set to $I=6,000$ in all simulation settings and we sampled a subset of $1,000$ points in each case. The data were centered around $(0,0,0)$, which was assumed known. \

{\bf Case 1} The data points in the first simulation case are uniformly distributed around a cylinder with an open seam. Set $\theta_i\sim U(0,2\pi-0.5)$, $\epsilon_i\sim N(0, 0.15^2)$ and $z_i\sim U(-3,3)$ i.i.d., then let
\begin{equation*}
\mathbf{x}_i=\left[\begin{array}{c}\cos{\theta_i}(1+\epsilon_i)\\ \sin{\theta_i}(1+\epsilon_i)\\z_i\end{array}\right].
\end{equation*}\

{\bf Case 2} The second case is based on the Himmelblau's function \cite{himmelblau1972applied}, which is $f(x, y) = (x^2+y-11)^2 + (x+y^2-7)^2 $. Let $z_{i1}\sim U(-5,5)$, $z_{i2}\sim U(-5,5)$, $\epsilon_i\sim N(0, 50^2)$ i.i.d., then let
\begin{equation*}
\mathbf{x}_i=\left[\begin{array}{c}z_{i1}\\ z_{i2}\\ -\frac{1}{100}\big\{(z_{i1}^2+z_{i2}-11)^2+(z_{i1}+z_{i2}^2-7)^2+\epsilon_i\big\}\end{array}\right].
\end{equation*}\

{\bf Case 3} In the third case the data points form a flatten surface at the beginning and then begin to bent over towards the bottom. Set $z_{i1}\sim U(0,2)$, $z_{i3}=0$ when $i=1,\dots,I/2$; $z_{i1}=\cos(\theta_i)+2$, $z_{i3}=-1+\sin(\theta_i)$ for $i=I/2+1,\dots,I$, where $\theta_i\sim U(-\pi/2,\pi/2)$. Let $z_{i2}\sim U(0,10)$ and $\epsilon_i\sim U(-0.4,0.4)$. All the random numbers are generated independently. Then we let
\begin{equation*}
\mathbf{x}_i=\left[\begin{array}{c}z_{i1}\\z_{i2}\\z_{i3}+\epsilon_i\end{array}\right].
\end{equation*}\

{\bf Case 4} In the last case, we produce a two dimensional stretched digit ``5". Let $z_{i1}\sim U(0,1)$, $z_{i3}=0$ when $i=1\dots,3I/10$; $z_{i1}=0$, $z_{i3}\sim U(-1,0)$ when $i=3I/10+1,\dots,4.5I/10$; $z_{i1}\sim U(0,0.5)$, $z_{i3}=-1$ when $i=4.5I/10+1,\dots,6I/10$; $z_{i1}=\frac{1}{2}+\frac{1}{2}\cos(\theta_i)$, $z_{i3}=-\frac{3}{2}+\frac{1}{2}\sin(\theta_i)$ for $i=6I/10+1,\dots,8.5I/10$ where $\theta_i\sim U(-\pi/2,\pi/2)$; $z_{i1}\sim U(0,0.5)$, $z_{i3}=-2$ when $i=8.5I/10+1,\dots,I$. Let $z_{i2}\sim U(0, 5)$ and $\epsilon_{i}\sim U(-0.15,0.15)$ when $i=1,\dots, I$. All the random numbers are generated independently. Then we let
\begin{equation*}
\mathbf{x}_i=\left[\begin{array}{c}z_{i1}+\epsilon_{i}\\z_{i2}\\z_{i3}+\epsilon_i\end{array}\right].
\end{equation*}\

\subsection{Simulation Results}
\begin{figure}[H]
\caption[Simulation results]{\footnotesize Simulation results. In the four top panels, the red-green dots are the original data points and the pink yellow ones are the fitted principal surfaces. The top panels from left to right show the results of (1) an cylinder with an open seam on one side, (2) the Himmelblau's function, (3) a non-function shaped ``carpet'' data cloud and (4) a simulated digit ``5'' data cloud. The four bottom panels compare the fitted surfaces with the original surfaces that generate the data cloud. The blue dots are the original surfaces and the pink yellow ones are the fitted principal surfaces.}
\label{fig.sim}
\begin{minipage}[b]{0.25\linewidth}
\centering
\scalebox{0.37}{\includegraphics{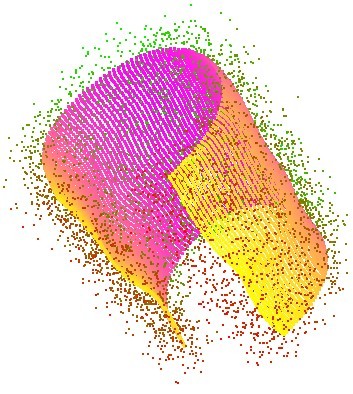}}
\scalebox{0.55}{\includegraphics{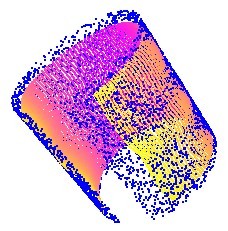}}
\end{minipage}
\hspace{-.4cm}
\begin{minipage}[b]{0.25\linewidth}
\centering
\scalebox{0.35}{\includegraphics{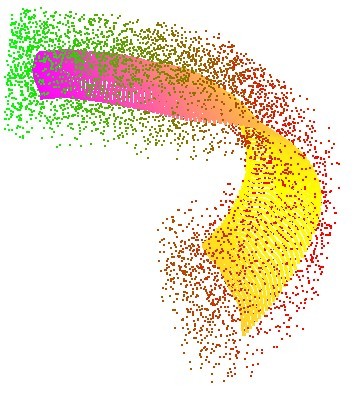}}
\scalebox{0.55}{\includegraphics{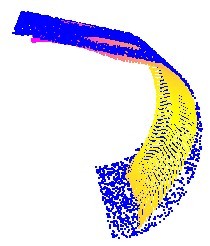}}
\end{minipage}
\hspace{-.3cm}
\begin{minipage}[b]{0.25\linewidth}
\centering
\scalebox{0.35}{\includegraphics{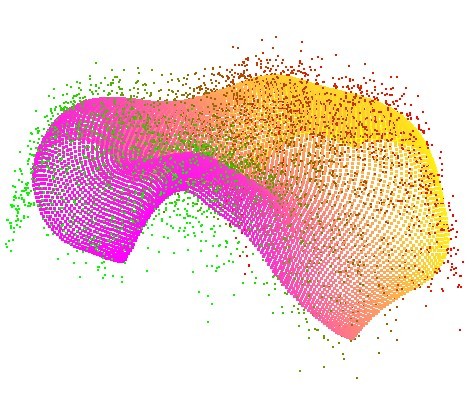}}
\scalebox{0.55}{\includegraphics{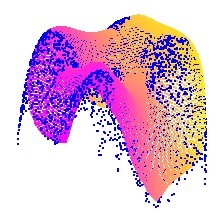}}
\end{minipage}
\hspace{-.1cm}
\begin{minipage}[b]{0.25\linewidth}
\centering
\vspace{-.2cm}
\scalebox{0.4}{\includegraphics{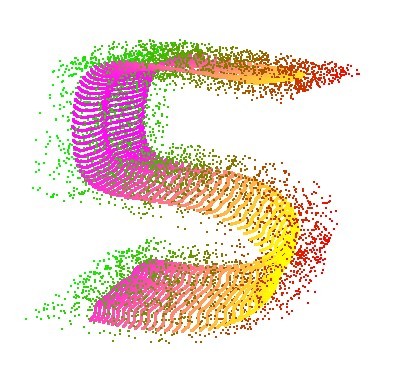}}
\scalebox{0.55}{\includegraphics{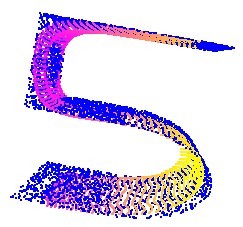}}
\end{minipage}
\end{figure}

The results of the principal surface fitting algorithm for four examples are shown in Figure \ref{fig.sim}. Starting in the upper left panel in Figure \ref{fig.sim}, one can see that the algorithm reconstructs the cylinder very well. Since we did not force any constraints in fitting the surface, we do not achieve a closed seam cylinder. (Finding principal surfaces for closed cylinders or spheres remains an interesting topic for future research.) Other upper panels illustrate the fitting results of the Himmelblau's function, a non-function shaped ``carpet'' data cloud and a simulated digit ``5'' data cloud. The four panels in the bottom panels in Figure \ref{fig.sim} compare the fitted surfaces with the original surfaces that generate the data cloud. The result are satisfactory for all cases. Given that the desired corpus callosum model fit is far simpler and smoother than these examples, the simulations show evidence of the viability of the principal surface algorithm as a robust method for this problem.\

For all four scenarios, the algorithm converged in less than 20 steps and took under two minutes on a i3-2.3GHz PC machine with 4Gb RAM memory. To further investigate practical computing speed, we set varied sample sizes and number of grid points ($N_{grid}$) for the third case above and measured the iterations and compute time. The results are illustrated in Table {\ref{tab.comp}}. As the number of grid points or the number of data points increases, the computing time for each iteration step increases correspondingly, though did not go out the range of reasonable compute times. Note that, one would rarely use the entire dataset in large sample size cases. Our recommended strategy is to take random subsamples from the dataset and subsequently fit the surface using the subsample, if necessary, repeating the sub-sampling to investigate its impact on model fits. In addition, as the grid number increases, the number of steps to attain convergence decreases.

\section{Application}
\subsection{Fitting the Principal Surface of a Corpus Callosum}
The data contain 466 scans generated from a diffusion tensor imaging (DTI) experiment performed on 176 MS patients. For each scan, we have corresponding fractional anisotropy (FA) value for each voxel of the entire corpus callosum area extracted via tractography. More detailed data description are provided in \cite{reich2010automated} and \cite{ozturk2010mri}. Fractional anisotropy has been associated cross sectionally and longitudinally with multiple sclerosis diagnoses and symptoms \cite{goldsmith2010penalized}, \cite{goldsmith2012longitudinal}, \cite{greven2011longitudinal}, \cite{reich2010automated}, \cite{zipunnikov2011functional}. Ignoring the FA value for the moment, we first discuss the principal surface fit.\

For each scan, there are roughly 20,000 data points in the corpus callosum region of interest. For computational simplicity, we randomly choose 1,000 of them to build the surface. We illustrate the result of four scans as well as the overall average fitted surface in the left panels in Figure \ref{fig.app}. The fitted surface for each scan is reasonable and is indicative of the fits from the other scans, each inspected visually. The overall average is smoother than each scan as we expected.

\subsection{Flattened FA Representation}
\begin{figure}[H]
\caption[Corpus callosum application results]{{\footnotesize In the left column, top four panels illustrates the the original corpus callosum from four different scans with their fitted principal surfaces. The scattered points are the original data points and the surface across the middle of them is the fitted principal surface. The bottom panel shows the average fitted principal surface for all the 466 scans. Red color suggests higher FA value while green means lower one. The right panels are the corresponding flattened surfaces. In the 2-D image, red color represents higher FA while green suggests lower one.}}
\label{fig.app}
\begin{minipage}[b]{0.5\linewidth}
\centering
\vspace{-1cm}
\scalebox{0.55}{\includegraphics{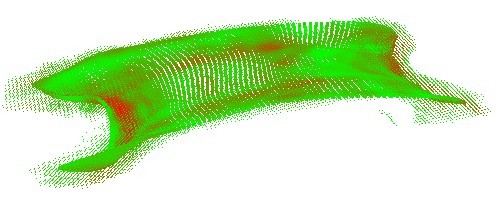}}
\vspace{-.1cm}
\scalebox{0.55}{\includegraphics{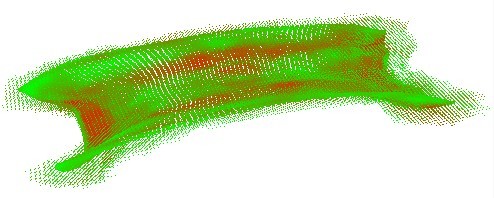}}
\vspace{-.1cm}
\scalebox{0.58}{\includegraphics{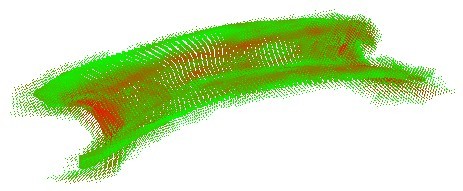}}
\vspace{-.1cm}
\scalebox{0.55}{\includegraphics{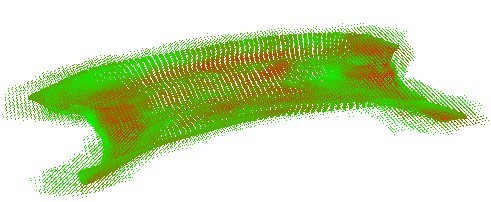}}
\vspace{-.1cm}
\scalebox{0.7}{\includegraphics{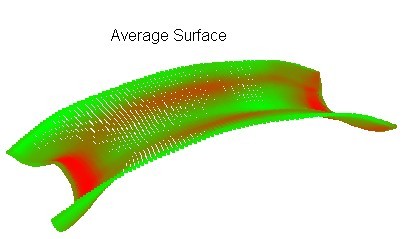}}
\end{minipage}
\begin{minipage}[b]{0.5\linewidth}
\vspace{-1cm}
\centering
\hspace{-0.6cm}
\scalebox{0.58}{\includegraphics{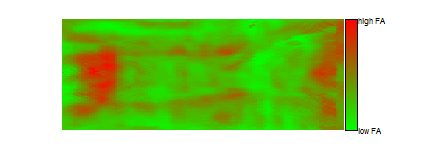}}
\vspace{-.15cm}
\scalebox{0.58}{\includegraphics{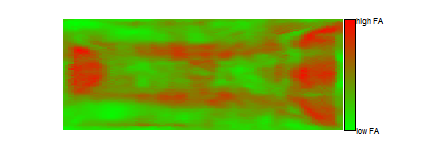}}
\vspace{-.15cm}
\scalebox{0.58}{\includegraphics{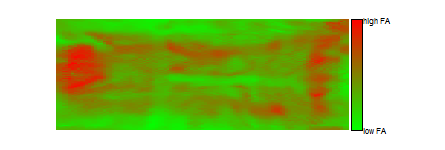}}
\vspace{-.15cm}
\scalebox{0.58}{\includegraphics{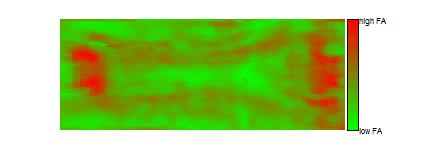}}
\vspace{-.15cm}
\hspace{0.4cm}
\scalebox{0.46}[0.53]{\includegraphics{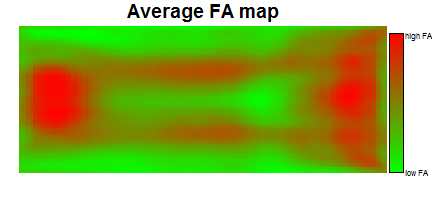}}
\end{minipage}
\end{figure}

A primary goal from the principal surface fitting is to flatten the surface and to make a 2D FA image. That is, our goal is to use DTI-based morphometric information to create 2D images of MR contrast properties, such as FA maps. First, we project each data point onto the surface by a grid search method. Then we smooth the associated FA value on the surface by calculating a local average. Interpolating this smooth onto a grid yields a $100\times 100$ 2D FA image displayed in the right panels of Figure \ref{fig.app}. That is, right panels represent the flattened surfaces of the corresponding left panels. The bottom panel shows the average FA map, which shows an overall visualization of 2D FA maps for all the scans. Another important fact is from left panels in Figure \ref{fig.app} and right panel in Figure \ref{fig.corpus}, we can see that important information of the FA values is retained, which is to say, it may be sufficient to visualize the 2D FA values instead of the original 3D FA values.

\section{Discussion}

In this manuscript we introduced a principal surface algorithm and used it to fit the corpus callosum. The goal of this work is largely developmental, creating a handy tool for
dimension reduction in DTI analysis of a primary structure. The result of both the simulations and the application is encouraging.
While applied to the corpus callosum, the algorithm could be applied to any other three dimensional manifold-like objects where a two dimensional surface could be embedded and is
of interest, for example, all the major white matter tracts in the brain \cite{bazin2011direct}. The two dimensional manifold characterizes the original data and accomplishes both dimension reduction and better visualization. The surface that we constructed is
smooth and could be easily projected onto to represent other properties of the original structure, such as the FA, mean diffusivity, parallel diffusivity, local thickness and so on. The algorithm
is computationally feasible and scales well to larger images and densely measured structures.\

The role of the two dimensional sub-representation needs to be further explored. For future work, we are developing functional data analysis tools  \cite{goldsmith2010penalized}, \cite{goldsmith2011functional}, \cite{greven2011longitudinal}, \cite{zhu2011fadtts}, \cite{zhu2010frats} for relating the dimension reduced 2D manifold to outcomes of interest for the purpose of inference, biomarker creation and prediction.
Of note, we are particularly interested in whether or not the 2D
representation of the corpus callosum is less sensitive to issues of whole brain registration often used in the processing pipeline. In fact, it is possible that
registering the 2D representation is preferable to whole brain registration a priori in certain applications. We note also that MFPCA \cite{di2009multilevel} and LFPCA \cite{zipunnikov2011functional} methods have been shown to isolate registration error
as a part of the model \cite{lee2012statistical}, thus raising the intriguing possibility of DTI processing streams that dramatically decrease the need and importance
of whole brain template-based registration.\

Furthermore, current works suggest sagittal mid-line callosum average thickness might be of very meaningful in the study of some diseases \cite{Luders2009decreased}, \cite{vidal2006mapping}. In contrast, The principal surface approach allows us to calculate the local thickness over the main body of the structure. We then could construct a global thickness map for the whole corpus callosum, following a procedure similar to that of constructing the FA map. Based on this, the thickness changes of white matter could be measured on the whole corpus callosum instead of the mid-sagittal slice.

Finally, we note that the procedure that we proposed can be further extended as a general problem of fitting skeleton manifolds.
First, we are interested in fitting surfaces with fixed boundaries. Work has been done to analyze principal curves with
fixed starting and ending points \cite{caffo2007acasestudy}. The extension to surfaces would be challenging. Suppose $\b{t}=(t_1, t_2)$ is the corresponding coordinate on the
surface of the original data point, $\b{x}=(x_1, x_2, x_3)$, which lies in 3D space. We restrict the range of $\b{t}$ to be $[0,1]\times[0,1]$. Now consider a predetermined
function for $\b{x}(\b{t})=(x_1(\b{t}), x_2(\b{t}),x_3(\b{t}))$ when the boundary values of $t$ are linearly constrained. Such constraints would yield cylindrical fits easily
though extensions to completely closed surfaces would require more elaborate constraints.

\section{Acknowledgements}
The project described was supported by Grant Number R01EB012547 and
NIH grant P41 EB015909 from the National Institute of Biomedical
Imaging And Bioengineering.  The project described was also supported by
Grant Number R01NS060910 from the National Institute of Neurological
Disorders and Stroke. Data for this study was acquired with grants from the National Multiple Sclerosis Society and EMD Serono.

\newpage
\begin{appendix}

\listoffigures
\listoftables

\newpage
\section{Figures}
\vspace{3cm}

\newpage

\smallskip

\newpage

\vspace{2cm}

\newpage

\smallskip

\newpage

\newpage

\section{Tables}

\begin{table}[H]
\caption[Computing time]{Simulation computing time consuming comparison}
\label{tab.comp}
\begin{minipage}[ht]{0.7\linewidth}
\centering
\begin{tabular}{|r|cccc|cccc|cccc|}
\hline
Sample & \multicolumn{12}{|c|}{Number of grid points, denoted as $N_{grid}$}\\
\cline{2-13}
&15 & 30 & 50 & 100  &  15 & 30 & 50 & 100  &  15 & 30 & 50 & 100  \\
\cline{2-13}
Size & \multicolumn{4}{|c}{Total time usage (Sec.)} & \multicolumn{4}{|c}{Number of iterations} & \multicolumn{4}{|c|}{Time per iteration (Sec.)}\\
\hline
100& 4 & 4 & 4 & 8  &  6  &  5  &  3  &  3 & 1 & 1 & 1 & 3\\
200&7 & 5 & 5 & 10  & 11  &  6  & 4  &  3 & 1 & 1 & 1 & 3\\
500&18 &12 &11 & 20  & 17  & 8  & 5  &  4 & 1 &1 &2 & 5\\
700&27& 21 &18 &32 &  20 &  12 &  7 &   5 &1& 2& 3& 6\\
1000&39& 40 &35& 49 &  22 &  15 &  10 &  6& 2& 3& 3& 8\\
\hline
\end{tabular}
\end{minipage}
\end{table}

\newpage

\bibliographystyle{plainnat}

\begin{thebibliography}{35}
\providecommand{\natexlab}[1]{#1}
\providecommand{\url}[1]{\texttt{#1}}
\expandafter\ifx\csname urlstyle\endcsname\relax
  \providecommand{\doi}[1]{doi: #1}\else
  \providecommand{\doi}{doi: \begingroup \urlstyle{rm}\Url}\fi

\bibitem[Bazin et~al.(2011)Bazin, Ye, Bogovic, Shiee, Reich, Prince, and
  Pham]{bazin2011direct}
Pierre-Louis Bazin, Chuyang Ye, John~A. Bogovic, Navid Shiee, Daniel~S. Reich,
  Jerry~L. Prince, and Dzung~L. Pham.
\newblock Direct segmentation of the major white matter tracts in diffusion
  tensor images.
\newblock \emph{NeuroImage}, 58\penalty0 (2):\penalty0 458 -- 468, 2011.

\bibitem[Caffo et~al.(2007)Caffo, Crainiceanu, Deng, and
  Hendrix]{caffo2007acasestudy}
Brian~S. Caffo, Ciprian~M. Crainiceanu, Lijuan Deng, and Craig~W. Hendrix.
\newblock A case study in pharmacologic imaging using principal curves in
  single photon emission computed tomography.
\newblock \emph{Johns Hopkins University, Dept. of Biostatistics Working
  Papers}, 143, 2007.

\bibitem[Di et~al.(2009)Di, Crainiceanu, Caffo, and Punjabi]{di2009multilevel}
C.~Di, C.~Crainiceanu, B.~Caffo, and N.~Punjabi.
\newblock Multilevel functional principal component analysis.
\newblock \emph{Annals of Applied Statistics}, 3\penalty0 (1):\penalty0
  458--488, 2009.

\bibitem[Dong and McAvoy(1996)]{dong1996nonlinear}
D.~Dong and T.J. McAvoy.
\newblock Nonlinear principal component analysis based on principal curves and
  neural networks.
\newblock \emph{Computers Chemical Engineering}, 20\penalty0 (1):\penalty0 65
  -- 78, 1996.
\newblock ISSN 0098-1354.

\bibitem[Duchon(1977)]{duchon1977splines}
J.~Duchon.
\newblock Splines minimizing rotation-invariant semi-norms in soboley spaces.
\newblock \emph{Lecture notes in mathematics}, 571/1977:\penalty0 85--100,
  1977.

\bibitem[Einbeck et~al.(2010)Einbeck, Evers, and Powell]{einbech2010data}
J.~Einbeck, L.~Evers, and B.~Powell.
\newblock Data compression and regression through local principal curves and
  surfaces.
\newblock \emph{International journal of neural systems}, 20\penalty0
  (3):\penalty0 177--192, 2010.

\bibitem[Fletcher(2004)]{fletcher2004statistical}
P.T. Fletcher.
\newblock Statistical variability in nonlinear spaces: Application to shape
  analysis and dt-mri phd thesis, university of north carolina at chapel hill.
\newblock 2004.

\bibitem[Fletcher et~al.(2004)Fletcher, Lu, Pizer, and
  Joshi]{fletcher2004medical}
P.T. Fletcher, Conglin Lu, S.M. Pizer, and Sarang Joshi.
\newblock Principal geodesic analysis for the study of nonlinear statistics of
  shape.
\newblock \emph{Medical Imaging, IEEE Transactions on}, 23\penalty0
  (8):\penalty0 995 --1005, aug. 2004.
\newblock ISSN 0278-0062.

\bibitem[Gerber et~al.(2009)Gerber, Tasdien, and
  Whitaker]{gerber2009dimensionality}
S.~Gerber, T.~Tasdien, and R.~Whitaker.
\newblock Dimensionality reduction and principal surfaces via kernel map
  manifolds.
\newblock \emph{Computer Vision, 2009 IEEE 12th International Conference},
  pages 529--536, 2009.

\bibitem[Gnanadesikan(1997)]{gnanadesikan1977methods}
H~Gnanadesikan.
\newblock \emph{Methods for Statistical Analysis of Multivariate Observations}.
\newblock Wiley, New York, 1997.

\bibitem[Goldsmith et~al.(2010)Goldsmith, Feder, Crainiceanu, Caffo, and
  Reich]{goldsmith2010penalized}
J.~Goldsmith, J.~Feder, C.M. Crainiceanu, B.~Caffo, and D.~Reich.
\newblock Penalized functional regression.
\newblock \emph{Johns Hopkins University, Dept. of Biostatistics Working
  Papers}, page 204, 2010.

\bibitem[Goldsmith et~al.(2011{\natexlab{a}})Goldsmith, Caffo, Crainiceanu,
  Reich, Du, and Hendrix]{goldsmith2011nonlinear}
J.~Goldsmith, B.~Caffo, C.~Crainiceanu, D.~Reich, Y.~Du, and C.~Hendrix.
\newblock Nonlinear tube-fitting for the analysis of anatomical and functional
  structures.
\newblock \emph{The annals of applied statistics}, 5\penalty0 (1):\penalty0
  337--363, 2011{\natexlab{a}}.

\bibitem[Goldsmith et~al.(2011{\natexlab{b}})Goldsmith, Crainiceanu, Caffo, and
  Reich]{goldsmith2011penalized1}
J.~Goldsmith, C.M. Crainiceanu, B.S. Caffo, and D.S. Reich.
\newblock Penalized functional regression analysis of white-matter tract
  profiles in multiple sclerosis.
\newblock \emph{NeuroImage}, 2011{\natexlab{b}}.

\bibitem[Goldsmith et~al.(2011{\natexlab{c}})Goldsmith, Wand, and
  Crainiceanu]{goldsmith2011functional}
J.~Goldsmith, M.P. Wand, and C.~Crainiceanu.
\newblock Functional regression via variational bayes.
\newblock \emph{Electronic Journal of Statistics}, 5:\penalty0 572--602,
  2011{\natexlab{c}}.

\bibitem[Goldsmith et~al.(2012)Goldsmith, Crainiceanu, Caffo, and
  Reich]{goldsmith2012longitudinal}
J.~Goldsmith, C.M. Crainiceanu, B.~Caffo, and D.~Reich.
\newblock Longitudinal penalized functional regression for cognitive outcomes
  on neuronal tract measurements.
\newblock \emph{Journal of the Royal Statistical Society: Series C (Applied
  Statistics)}, 2012.

\bibitem[Greven et~al.(2011)Greven, Crainiceanu, Caffo, and
  Reich]{greven2011longitudinal}
S.~Greven, C.~Crainiceanu, B.~Caffo, and D.~Reich.
\newblock Longitudinal functional principal component analysis.
\newblock \emph{Recent Advances in Functional Data Analysis and Related
  Topics}, pages 149--154, 2011.

\bibitem[Hastie(1984)]{hastie1984principal}
T.~Hastie.
\newblock Principal curves and surfaces.
\newblock \emph{Technical report}, 11, 1984.

\bibitem[Hastie and Stuetzle(1989)]{hastie1989principal}
T.~Hastie and W.~Stuetzle.
\newblock Principal curves.
\newblock \emph{Journal of the American Statistical Association}, 84\penalty0
  (406):\penalty0 502--516, 1989.

\bibitem[Himmelblau(1972)]{himmelblau1972applied}
David~M. Himmelblau.
\newblock \emph{Applied nonlinear programming}.
\newblock McGraw-Hill (New York), 1972.

\bibitem[Kohonen(1990)]{kohunen1990theself}
T.~Kohonen.
\newblock The self-organizing map.
\newblock \emph{Proceedings of the IEEE}, 78\penalty0 (9):\penalty0 1464
  --1480, sep 1990.
\newblock ISSN 0018-9219.

\bibitem[Kohonen(1982)]{kohonen1982self-organize}
Teuvo Kohonen.
\newblock Self-organized formation of topologically correct feature maps.
\newblock \emph{Biological Cybernetics}, 43:\penalty0 59--69, 1982.
\newblock ISSN 0340-1200.

\bibitem[Kramer(1991)]{kramer1991nonlinear}
Mark~A. Kramer.
\newblock Nonlinear principal component analysis using autoassociative neural
  networks.
\newblock \emph{AIChE Journal}, 37\penalty0 (2):\penalty0 233--243, 1991.
\newblock ISSN 1547-5905.

\bibitem[Kramer(1992)]{kramer1992autoassociative}
Mark~A. Kramer.
\newblock Autoassociative neural networks.
\newblock \emph{Computers and Chemical Engineering}, 16\penalty0 (4):\penalty0
  313 -- 328, 1992.
\newblock ISSN 0098-1354.
\newblock Neutral network applications in chemical engineering.

\bibitem[Leblanc and Tibshirani(1994)]{leblanc1994adaptive}
Michael Leblanc and Robert Tibshirani.
\newblock Adaptive principal surfaces.
\newblock \emph{Journal of the American Statistical Association}, 89\penalty0
  (425):\penalty0 53--64, 1994.

\bibitem[Lee et~al.(2012)Lee, Zipunnikov, Caffo, Reich, and
  Pham]{lee2012statistical}
S~Lee, V~Zipunnikov, B~Caffo, D~Reich, and D~Pham.
\newblock Statistical image analysis of longitudinal ravens images: methodology
  and case study.
\newblock \emph{Technical Report}, 2012.

\bibitem[Luders et~al.(2009)Luders, Narr, Hamilton, Phillips, Thompson, Valle,
  Del'Homme, Strickland, McCracken, Toga, and Levitt]{Luders2009decreased}
Eileen Luders, Katherine~L. Narr, Liberty~S. Hamilton, Owen~R. Phillips,
  Paul~M. Thompson, Jessica~S. Valle, Melissa Del'Homme, Tony Strickland,
  James~T. McCracken, Arthur~W. Toga, and Jennifer~G. Levitt.
\newblock Decreased callosal thickness in attention-deficit/hyperactivity
  disorder.
\newblock \emph{Biological Psychiatry}, 65\penalty0 (1):\penalty0 84 -- 88,
  2009.

\bibitem[Ozertem and Erdogmus(2011)]{ozertem2011locally}
U.~Ozertem and D.~Erdogmus.
\newblock Locally defined principal curves and surfaces.
\newblock \emph{Journal of Machine learning research}, 12:\penalty0 1249--1286,
  2011.

\bibitem[Ozturk et~al.(2010)Ozturk, Smith, Gordon-Lipkin, Harrison, Shiee,
  Pham, Caffo, Calabresi, and Reich]{ozturk2010mri}
A.~Ozturk, SA~Smith, EM~Gordon-Lipkin, DM~Harrison, N.~Shiee, DL~Pham,
  BS~Caffo, PA~Calabresi, and DS~Reich.
\newblock Mri of the corpus callosum in multiple sclerosis: association with
  disability.
\newblock \emph{Multiple Sclerosis}, 16\penalty0 (2):\penalty0 166--177, 2010.

\bibitem[Palus and Dvorak(1992)]{palus1992singular}
Milan Palus and Ivan Dvorak.
\newblock Singular-value decomposition in attractor reconstruction: Pitfalls
  and precautions.
\newblock \emph{Physica D: Nonlinear Phenomena}, 55\penalty0 (1–2):\penalty0
  221 -- 234, 1992.
\newblock ISSN 0167-2789.

\bibitem[Reich et~al.(2010)Reich, Ozturk, Calabresi, and
  Mori]{reich2010automated}
D.~Reich, A.~Ozturk, P.~Calabresi, and S.~Mori.
\newblock Automated vs. conventional tractography in multiple sclerosis:
  Variability and correlation with disability.
\newblock \emph{NeuroImage}, 49:\penalty0 3047--3056, 2010.

\bibitem[Vidal et~al.(2006)Vidal, Nicolson, DeVito, Hayashi, Geaga, Drost,
  Williamson, Rajakumar, Sui, Dutton, et~al.]{vidal2006mapping}
C.N. Vidal, R.~Nicolson, T.J. DeVito, K.M. Hayashi, J.A. Geaga, D.J. Drost,
  P.C. Williamson, N.~Rajakumar, Y.~Sui, R.A. Dutton, et~al.
\newblock Mapping corpus callosum deficits in autism: an index of aberrant
  cortical connectivity.
\newblock \emph{Biological Psychiatry}, 60\penalty0 (3):\penalty0 218--225,
  2006.

\bibitem[Wood(2003)]{wood2003thin}
S.~Wood.
\newblock Thin plate regression splines.
\newblock \emph{Journal of the royal statistical society}, 65\penalty0
  (1):\penalty0 95--114, 2003.

\bibitem[Zhu et~al.(2010)Zhu, Styner, Tang, Liu, Lin, and
  Gilmore]{zhu2010frats}
H.~Zhu, M.~Styner, N.~Tang, Z.~Liu, W.~Lin, and J.H. Gilmore.
\newblock Frats: Functional regression analysis of dti tract statistics.
\newblock \emph{Medical Imaging, IEEE Transactions on}, 29\penalty0
  (4):\penalty0 1039--1049, 2010.

\bibitem[Zhu et~al.(2011)Zhu, Kong, Li, Styner, Gerig, Lin, and
  Gilmore]{zhu2011fadtts}
H.~Zhu, L.~Kong, R.~Li, M.~Styner, G.~Gerig, W.~Lin, and J.H. Gilmore.
\newblock Fadtts: Functional analysis of diffusion tensor tract statistics.
\newblock \emph{NeuroImage}, 56\penalty0 (3):\penalty0 1412--1425, 2011.

\bibitem[Zipunnikov et~al.(2011)Zipunnikov, Greven, Caffo, Reich, and
  Crainiceanu]{zipunnikov2011functional}
Vadim Zipunnikov, Sonja Greven, Brian Caffo, Daniel~S. Reich, and Ciprian
  Crainiceanu.
\newblock Longitudinal high-dimensional data analysis.
\newblock \emph{Johns Hopkins University, Dept. of Biostatistics Working
  Papers}, 234, 2011.

\end{thebibliography}

\end{appendix}
\end{document}